\documentclass[nonacm, sigchi]{acmart}
\renewcommand\footnotetextcopyrightpermission[1]{} 
\usepackage{float}
\usepackage{hyperref}
\usepackage{multicol}
\usepackage{lipsum}
\usepackage{graphicx}
\AtBeginDocument{%
  \providecommand\BibTeX{{%
    \normalfont B\kern-0.5em{\scshape i\kern-0.25em b}\kern-0.8em\TeX}}}

\begin{document}

\title{Flat combined Red Black Trees}

\author{Sergio Sainz}
\email{ssainz@vt.edu}
\affiliation{%
  \institution{Virginia Polytechnic}
  \city{Blacksburg}
  \state{Virginia}
}

\begin{abstract}
Flat combining is a concurrency threaded technique whereby one thread performs all the operations in batch by scanning a queue of operations to-be-done and performing them together. Flat combining makes sense as long as k operations each taking O(n) separately can be batched together and done in less than O(k*n). Red black tree is a balanced binary search tree with permanent balancing warranties. Operations in red black tree are hard to batch together: for example inserting nodes in two different branches of the tree affect different areas of the tree. In this paper we investigate alternatives to making a flat combine approach work for red black trees.
\end{abstract}



\maketitle

\section{Introduction}
Flat combining is a concurrent programming technique whereby one thread performs all the operations in batch by scanning a queue of operations to-be-done and performing them together meanwhile caller threads wait in queue. Flat combining makes sense as long as k operations each taking O(n) separately can be batched together and done in less than O(k*n). Red black tree is a balanced binary search tree with permanent balanced tree warranties. Operations in red black tree are hard to batch together: for example inserting nodes in two different branches of the tree affect different areas of the tree and thus cannot be done together in a cheaper, obvious way. Thus, this is the problem: how to achieve fast concurrent red black tree using flat combining technique if operations cannot be batched together in an obvious manner.
\\
Flat combining is introduced by Hendler, Incze, Shavit, and Tzafrir (\cite{FlatCombine}). The idea of how it works is as follows: (1) Each thread allocates its operation in a queue and (2) thread checks if structure lock is free, if so, take it and scan the queue for pending operations. (3) perform operations in batch. (4) return lock. 

\begin{figure}[H]
  \includegraphics[width=\linewidth]{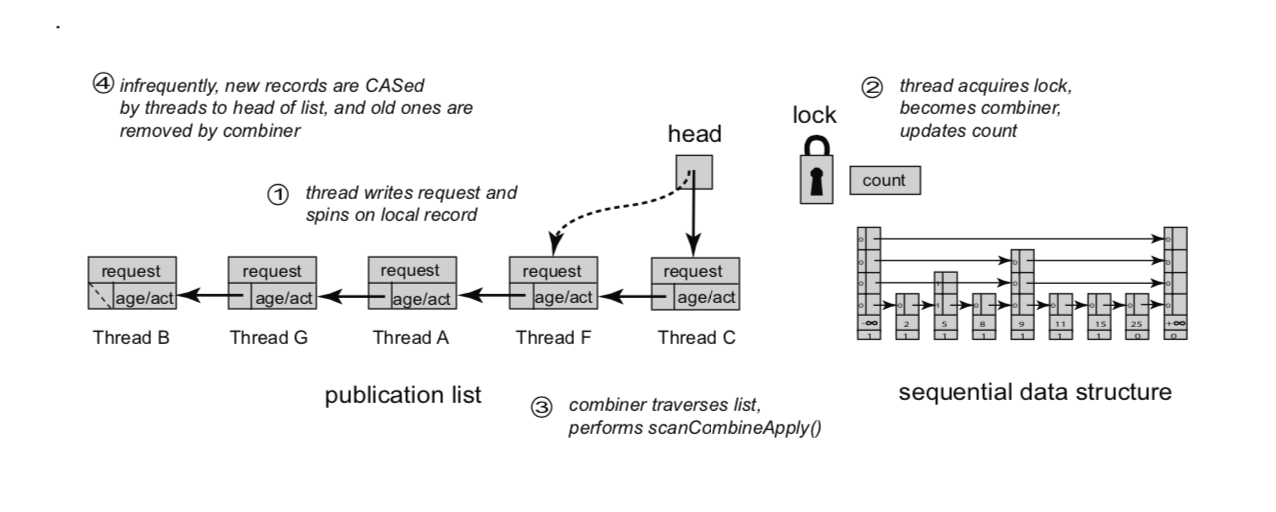}
  \caption{Flat combine explanation diagram}
  \label{fig:boat1}
\end{figure}

Proposed red black tree has differences with original flat combine approach in two main areas:

\begin{enumerate}
    \item Flat combine red black tree uses only one thread as the combiner . Meanwhile, in work of \cite{FlatCombine} , the thread that acquires lock takes responsibility of being combiner. 
    \item Flat combine red black tree let caller thread perform some of the operations (such as GET). Meanwhile, in work of \cite{FlatCombine}, the combiner is always performing the operations by itself. 
\end{enumerate}

\section{Previous work}
There is no previous work exploring flat combining and red black trees. There is also little to no references about implemented lock-free red black trees. Work by Kim, Cameron and Graham \cite{LFRBT} explains that there is a lock free red black tree but the implementation details did not explain how to implement the method moveUpStruct. Proposed method is benchmarked against two other concurrent approaches, albeit are not lock-free: (1) Compositional red black tree \url{https://github.com/gramoli/synchrobench/blob/master/java/src/trees/transactional/CompositionalRBTreeIntSet.java} and (2) Transactional red black tree. \url{https://github.com/gramoli/synchrobench/blob/master/java/src/trees/transactional/TransactionalRBTreeSet.java }. Author for both methods is Maurice Herlihy (author of Art of Multiprocessor Programming book \cite{ArtOfMultiprocessor}).

\section{Scope}

In the experiments we use the load of : 10\% insert operations, 10\% delete operations, and 80\% get operations. The benchmarks are the compositional red black tree and the transactional red black tree as mentioned earlier. And also a coarsed grained red black tree. No AVL or B-Tree are used for the benchmark although they do have available lock-free versions. I plan to add them if time permits but time was unavailable and hence the testing remains pending.

\section{Solutions}

I chose Java as programming language because is the language we studied the concurrency concepts on. Thus I am more familiar with it.

\subsection{Version 1: blocking queue and sleep/awake}

This solution consists of one thread dedicated to be combiner tasks. It starts with the tree and it waits for all other caller threads to add their operations into the operations queue. The queue is a blocking queue in the sense that once item is added to the queue , the caller waits until the operation is dequeued. Then, once the caller thread finishes enqueueing the operator item (because combiner thread already dequeues the operator item), caller thread will go and sleep meanwhile combiner analyses the operator item. Operator item could be three kinds: DELETE, INSERT or GET. 

\begin{enumerate}
    \item In case the operator is GET, the combiner will let the caller perform the GET operation and will increment an atomic counter, \textbf{numberOfPendingGet}, to know when the caller thread is finished with the GET (the caller will decrease \textbf{numberOfPendingGet}. 
    \item In case the operator is DELETE, the combiner will wait until \textbf{numberOfPendingGet} is 0 and if so, it will perform delete.
    \item In case the operator is INSERT, the combiner will wait until \textbf{numberOfPendingGet} is 0 and if so, it will perform insert.
\end{enumerate}

\begin{figure}[H]
  \includegraphics[width=\linewidth]{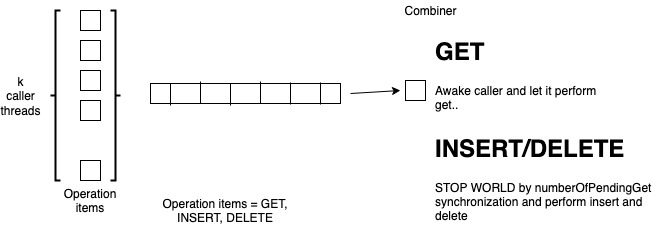}
  \caption{Version 1 diagram}
  \label{fig:boat1}
\end{figure}

\subsection{Version 2}

Version 2 performs the same as version 1, except that version 2 GET operations will not sleep and instead they will spin for the combiner on an atomic boolean variable.

\subsection{Version 3} 

Version 3 performs the same as version 2, except that in version 3 the DELETE And INSERT operations do not sleep after they finish enqueueing the operation item. Instead, they will spin in an atomic integer and use fixed backoff strategy to sleep intermittently. For example: If caller thread finishes enqueueing the operation item, it goes and check whether atomic boolean variable to continue is set, and if not, thread goes to sleep. The first time the thread finds the atomic boolean variable is not set it will go to sleep for 1 millisecond, then second time for 3 milliseconds and so on. The fixed sequence of sleep times is 1, 3, 10, 20, 50, 100, 200, 500, 1000, 3033, 5000 (these values are selected manually randomly increasing). This is the best backoff strategy obtained in the course homework related to backoff lock.

\subsection{Version 4}

Version 4 performs the same as version 2, except that in version 4 the DELETE and INSERT operations do not sleep after they finish enqueueing the operation item. Instead, they will spin in atomic integer continuously. 

\subsection{Version 5}

Version 5 introduces the notion of \textbf{soft delete} and \textbf{soft insert}. These two operations are inspired by the list deletion where nodes are marked to be deleted prior actual deletion. Both of these operations : soft delete and soft insert, will operate only if the key being inserted or deleted already exists in the tree. If that is the case, then an atomic boolean variable \textbf{isDeleted} will be turned on (if soft delete) or turned off (if soft insert).

\begin{figure}[H]
  \includegraphics[width=\linewidth]{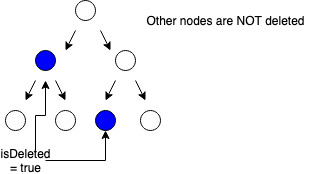}
  \caption{In blue are "soft deleted" nodes. In white are "not deleted" nodes.}
  \label{fig:boat1}
\end{figure}

Then, combiner has these different rules:
\begin{enumerate}
    \item In case the operator is GET, the combiner will let the caller perform the GET operation and will increment an atomic counter, \textbf{numberOfPendingOperations}, to know when the caller thread is finished with the GET (the caller will decrease \textbf{numberOfPendingOperations} once done with GET). 
    \item In case the operator is DELETE, the combiner will check if key already exists in tree and if so increment \textbf{numberOfPendingOperations} and let caller soft delete the key. Else, it will just awake the caller for caller to continue (as the key does not exists in tree).
    \item In case the operator is INSERT, the combiner will check if key exists in tree and if so increment \textbf{numberOfPendingOperations} by one and let caller do soft insert. Otherwise if the key does not exists, combiner STOPS WORLD by turning on flag stopWorld. Then the combiner will wait until all threads will wait until \textbf{numberOfPendingOperations} is 0 and if so, it will perform insert. In case the number of nodes in tree is greater than the MAX number of nodes in tree, it will also actually delete the deleted nodes from the tree. 
    
\end{enumerate}

\subsection{Version 6}

Version 6 is same as version 5 but instead of using \textbf{numberOfPendingOperations} atomic variable to synchronize the operations it uses thread local variables. Also, in version 6 the GET operations do not enqueue to the queue. They just run by the caller directly in the tree but they do check their individual thread local Stop World atomic variables.

\subsection{Version 5 and Version 6 linearization}

Here it becomes tricky how to linearize the soft INSERT and soft DELETE when the concurrent GET operations occur. We assume first the soft INSERT operation is going on to insert previously marked as deleted node key 13, then some caller threads get key 13 as well. The caller thread performing the soft INSERT will at some point find the node whose key is 13, then before setting the node's \textbf{isDeleted} atomic boolean as false, some caller GET threads will read the node's previous \textbf{isDeleted} previous value and return. Once the INSERT thread marks the \textbf{isDeleted} as false, some other GET threads will read the \textbf{isDeleted} as false then return the actual Value for the node. Hence, linearization point for the soft insert is the moment the INSERT thread marks the \textbf{isDeleted} as false. Conversely, the same is true for soft DELETE operation. Because in both soft DELETE and soft INSERT the node already exists in the tree. 
\\
For actual insert, it happens in Stop World situations and thus only combiner is sequentially inserting and deleting (linearization points are not necessary for sequential insert/delete). 
\\
Please find below linearization points for soft delete and soft insert.
\begin{figure}[H]
  \includegraphics[width=\linewidth]{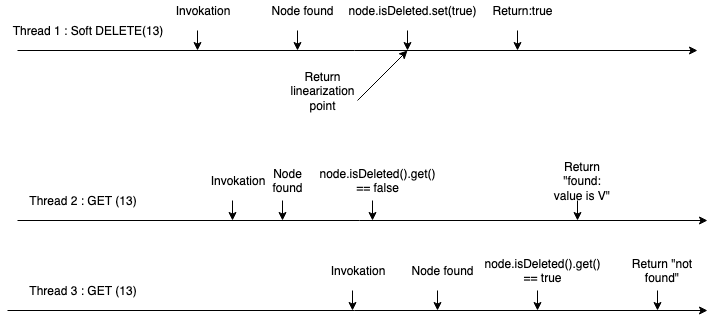}
  \caption{Soft delete linearization point}
  \label{fig:boat1}
\end{figure}
\begin{figure}[H]
  \includegraphics[width=\linewidth]{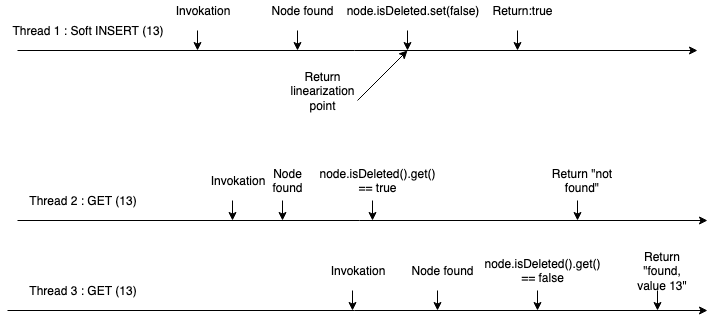}
  \caption{Soft insert linearization point}
  \label{fig:boat1}
\end{figure}

\subsection{Limitations and future work}

Each of the limitations below should be investigated.
\begin{enumerate}
    \item As of now the best performing tree, version 6 uses fixed number of threads. And each thread requires to register its thread local variable in the data structure prior any thread submitting requests.
    \item The memory of the tree determines its parallelism. The less memory the more frequent sequential Stop World operations will happen. Thus this is not suitable for environments without much memory.
    \item Also, the load assumes high number of collisions when getting, deleting and inserting. If the number of collisions is sparse, then there will be little use for the soft insert and thus there will again be more frequent sequential Stop World operations to actually remove the deleted nodes from the tree once they reach the max memory limit.
\end{enumerate}

\section{RESULTS}

\begin{figure*}[h]
  \includegraphics[width=\linewidth]{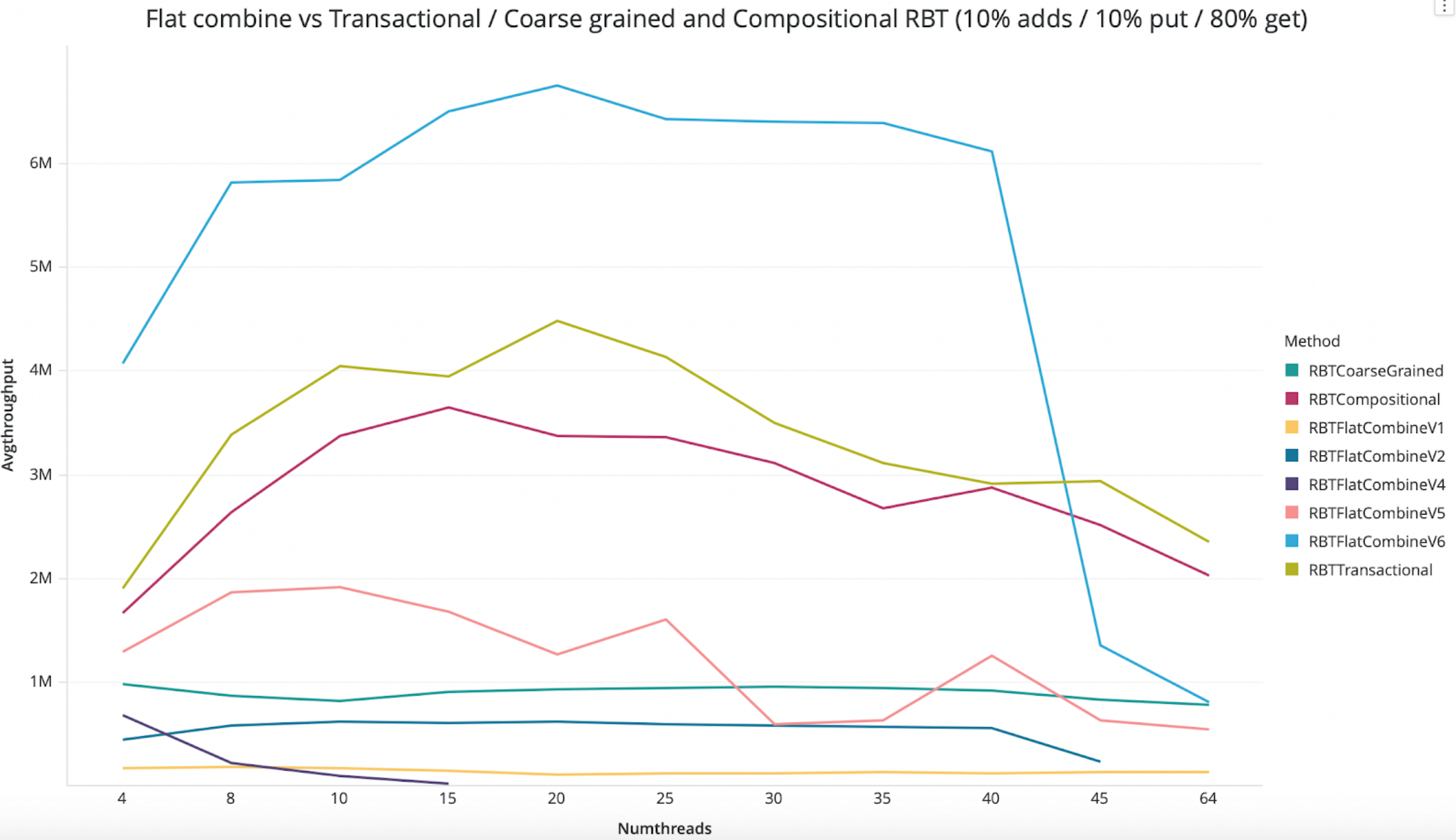}
  \caption{Results}
  \label{fig:results}
\end{figure*}

As mentioned earlier in the experiments we use the load of : 10\% insert operations, 10\% delete operations, and 80\% get operations. The benchmarks are the compositional red black tree and the transactional red black tree as mentioned earlier. And also a coarsed grained red black tree. Please find results in figure \ref{fig:results}. The max size of the tree is 1000 nodes and the range where the values are taken from to insert/delete/get is from 0 to 2000. 640000 iterations are submitted in each thread combination. 

\subsection{Conclusions}
\begin{enumerate}
    \item Versions 1 to 6 need TWO steps of synchronization. And this has big hits in performance when spinning is done in same variable (versions 1 - 5). Because all versions use blocking queue they need to wait for combiner to pick their item spinning and then further go and wait for combiner to complete its processing. Versions 1 to 6 use a blocking queue to linearize operations to combiner. All those versions need to spin until combiner picks their operation, then further wait either by going to sleep (versions 1 (get, delete, insert), 2 (delete, insert)) or spin (version 3 spins with backoff strategy, version 4 spins) or combination of both ( version 6 depends whether Stop the world is called and sleeps or not, then spins). Conclusion is to place in queue and go to sleep right away (instead of spinning). There is no time to test this hypothesis but it can be tried next as \textbf{future work}.
    \item Due to previous point, once the spinning stopped being in same variable, and instead done in each thread local atomic variable , the performance increased greatly (from version 5 to version 6).
    \item Also due to first point: Versions 1- 4 perform worse than coarse grained tree. This is because of double synchronizations. 
    \item Version 6 performs very bad after threads beyond 45. Reason for this is perhaps the multiple spinning going on when caller threads are waiting for the items in the queue to be picked up by combiner. Suggest in \textbf{future work} to use a non-blocking queue where the items are dropped and then goes to sleep directly afterwards (just using one synchronization point). 
\end{enumerate}

\bibliographystyle{ACM-Reference-Format}
\bibliography{biblio}

\end{document}